Superconductivity at 14K in the Co-doped $SmFe_{0.9}Co_{0.1}AsO$

V.P.S. Awana*, Arpita Vajpayee, Anand Pal, Monika Mudgel, R.S. Meena and H. Kishan

National Physical Laboratory, Dr. K.S. Krishnan Marg, New Delhi 110012, India

# ABSTRACT

We report superconductivity in the $SmFe_{0.9}Co_{0.1}AsO$ system being prepared by most easy and versatile single step solid-state reaction route. The parent compound SmFeAsO is non-superconducting but shows the spin density wave (*SDW*) like antiferromagnetic ordering at around 140K. To destroy the antiferromagnetic ordering and to induce the superconductivity in the parent system, the $Fe^{2+}$ is substituted partially by $Co^{3+}$. Superconductivity appears in $SmFe_{0.9}Co_{0.1}AsO$ system at around 14K. The Co doping suppresses the *SDW* anomaly in the parent compound and induces the superconductivity. Magnetization measurements show clearly the onset of superconductivity with $T_c^{dia}$ at 14K. The isothermal magnetization measurements exhibit the lower critical fields ($H_{c1}$) to be around 200Oe at 2 K. The bulk superconductivity of the studied $SmFe_{0.9}Co_{0.1}AsO$ sample is further established by open diamagnetic $M(H)$ loops at 2, and 5K. Normal state (above $T_c$) linear isothermal magnetization $M(H)$ plots excluded presence of any ordered magnetic impurity in the studied compound.

* Corresponding Author: Dr. V.P.S. Awana
Room 109,
National Physical Laboratory, Dr. K.S. Krishnan Marg, New Delhi-110012, India
Fax No. 0091-11-45609310: Phone no. 0091-11-45609210
e-mail-awana@mail.nplindia.ernet.in: www.freewebs.com/vpsawana/

## INTRODUCTION

The recent discovery of superconductivity at 26K in $LaFeAsO_{1-x}F_x$ and the subsequent findings of the enhanced $T_c$ up to 56K have attracted a great deal of attraction, in particular from the superconductors scientific community [1-5]. The parent compounds of this new superconductor ReFeAsO (Re = La, Pr, Sm, Gd, Nd, Ce etc.) have a quasi two-dimensional tetragonal layered structure. The Fe-As layer is usually thought to be responsible for superconductivity and the Re-O layer is a carrier reservoir layer to provide electron/hole carriers in ReFeAsO compounds. The parent compounds of this family do not show superconductivity at all and rather exhibits an anomaly in resistivity measurement attributed to spin density wave (*SDW*) characteristic [6,7]. The doping of electrons/hole carriers into the Fe-As layer leads to the suppression of the long range *SDW* order and to the subsequent induction of the superconductivity in these compounds [1-5,8-10]. The doping of electrons is achieved either by oxygen deficiency or by fluorine (F) doping [1-5,8,10]. The former can be achieved by high pressure high temperature (*HPHT*) only and later by either *HPHT* or quartz encapsulation coupled with high temperature heat treatment. Interestingly *HPHT* (6 Gpa, 1400 $^0$C) is not readily available in most laboratories and the fluorine corrodes with the quartz at high temperatures. Hence, in principle it is getting difficult for material scientists to synthesize these superconductors. Recently Sefat et al [11] and Qi et al [12] have come up with an idea to induce superconductivity in these Iron-Arsenide compounds by doping of electrons through $Fe^{2+}$ site $Co^{3+}$ substitution in the Fe-As layer. We followed this and achieved bulk superconductivity at about 14K in $SmFe_{0.9}Co_{0.1}AsO$ sample. The SDW characteristic of the SmFeAsO was suppressed by Co doping and superconductivity was induced at around 14K.

## EXPERIMENTAL

Polycrystalline SmFeAsO and $SmFe_{0.9}Co_{0.1}AsO$ samples were synthesized by single step solid-state reaction method. Stoichiometric amounts of Sm, Fe, As, $Co_3O_4$ and $Fe_2O_3$ were thoroughly ground. It is to be noted that weighing and grinding was done in the glove box under high purity argon atmosphere. The powder was then palletized and vacuum-sealed in a quartz tube. Roughly the vacuum was of the order of $10^{-4}$ Torr. Subsequently this sealed quartz ampoule was placed in box furnace and heat treated at 550°C for 12 hours, 850°C for 12 hours and then finally 1150°C for 33 hours in continuum. Then furnace was allowed to cool naturally. The sintered sample was obtained by breaking the quartz tube. The obtained sample is black in color and bit brittle, in fact it is nearly in powder form. Room temperature X-ray diffraction pattern was obtained using Cu$K_\alpha$ radiation on Desktop Rigaku-miniflex II diffractometer. The magnetization measurements were carried out on Quantum Design magnetic property measurement system (*MPMS*).

## RESULTS AND DISCUSSION

Figure 1 shows the room temperature *XRD* patterns for SmFeAsO and $SmFe_{0.9}Co_{0.1}AsO$ samples along with their Rietveld refinements. It is observed that all main peaks can be well indexed based on the basis of space group *P*4/*nmm*. Further, besides the majority phase (tetragonal *P4/nmm*) an extra peak at around 27.80 degree having low intensity (marked with *) is also seen in the XRD pattern of $SmFe_{0.9}Co_{0.1}AsO$ sample. All the permitted diffraction planes of the ReFeAsO system are marked with green vertical lines between the observed/fitted patterns and their difference in the bottom. It can be noted that except the * marked peak at 27.80 degree all other observed peaks match with the identified green marked planes of the system. The same impurity is seen for $SmFe_{0.9}Co_{0.1}AsO$ in ref. 12 as well. The structure of SmFeAsO and $SmFe_{0.9}Co_{0.1}AsO$ at 300 K is refined with the tetragonal space group *P*4/*nmm*. Sm and As atoms are located at Wyckoff positions *2c*, O is situated at *2a* and Fe/Co are shared at site *2b*. The Lattice parameters for $SmFe_{0.9}Co_{0.1}AsO$ are found to be $a$ = 3.9398 (7) Å and $c$ = 8.4639 (4) Å. While for the parent compound the lattice parameters are $a$ = 3.9375 (6) Å and $c$ = 8.5021 (4) Å. It is clear that the $c$ lattice parameter decreases with Co doping, while the $a$-axis remains nearly unaltered. The decrease in $c$ lattice parameter suggests that Co has gone in the tri-valence $Co^{3+}$ state at the place of $Fe^{2+}$, because ionic size of $Co^{3+}$ is smaller than $Fe^{2+}$. This substitution of $Co^{3+}$ at the site of $Fe^{2+}$ increases electron carriers in Fe-As layer, which in turn induces the superconductivity in Co doped SmFeAsO, is to be discussed next. The electrical and thermal characterization of the pristine SmFeAsO has already been reported by some of us in ref.7. This compound clearly exhibited a *SDW* character below 150K. Further in SmFeAsO, the $Sm^{3+}$ ions ordered anti-Ferro magnetically (*AFM*) and superconductivity was not observed down to 2K, for details see ref. 7.

The temperature dependence of magnetic susceptibility, measured in both zero-field-cooled (*ZFC*) and field-cooled (*FC*) conditions at 10Oe for $SmFe_{0.9}Co_{0.1}AsO$ is represented in figure 2. Although the transition is broad, both *ZFC* and *FC* clearly indicate the transition of the compound to a superconducting state below 14K. The magnetic signal comes out to be negative below 14K. The onset temperature of superconducting transition $T_c^{dia}$ for $SmFe_{0.9}Co_{0.1}AsO$ is hence considered at 14K. The magnitude of the magnetic signal confirms the bulk superconductivity in our sample. As far as the volume fraction of superconductivity is concerned, the signal magnitude is slightly better than as in ref. 11 for $LaFe_{0.89}Co_{0.11}AsO$ and in ref. 12 for $SmFe_{0.9}Co_{0.1}AsO$. Also as rightly pointed out in ref.11, the determination of superconducting volume fraction on sintered polycrystalline samples had always been tricky due to pinning and penetration effects. Never the less we can safely conclude that the presently studied single step synthesized $SmFe_{0.9}Co_{0.1}AsO$ sample is superconducting below 14K, and is in confirmation with the results shown in ref. 11 and ref. 12. Interestingly although the optimum superconducting transition temperature ($T_c$) for the doped (either by flourine or oxygen vacancy) LaFeAsO and SmFeAsO are around 30 and 55K respectively [1,5], with Co doping both are superconducting below about 14K [11,12]. In another very recent report on Co doped CeFeAsO, superconductivity is observed below 11.3K [13]. It is clear that in Co doped samples neither the optimum $Tc$ nor the obvious Re ionic size effect [1,5,14] are yet established [6,7,13]. We believe the possible reasoning behind the same could be disorder in superconducting Fe-As block due to $Fe^{2+}/Co^{3+}$ substitution. Although the $Co^{3+}$ substitution at $Fe^{2+}$ site in Fe-As layer provides the required mobile electrons but creates disorder as well in the superconductivity layer. It is also to be noted, that although bulk superconductivity in terms of diamagnetic transitions is seen in both ref. 11 and in presently studied sample, the volume fraction is seemingly less and can not be left without discussion, as suggested in ref.11. Though Co substitution at Fe site induces superconductivity in Fe-As based oxy-arsenides by electron doping, the same induces disorder and hence the superconductivity is weak and not optimized. Seemingly there is a competition between direct carrier introduced ($Fe^{2+}/Co^{3+}$) superconductivity and the disorder thus created. Or otherwise principally 10% substitution of $O^{-2}$ by $F^{-1}$ or oxygen vacancies of around 15% must yield the similar result with same amount of $Co^{3+}$ at $Fe^{2+}$ site. Another important possibility still to be probed is exact amount of doped $Co^{3+}$, some of the substituted Co may be in +2 state as well, and hence ineffective in doping electrons. In any case even by varying the Co content in $LaFe_{1-x}Co_xAsO$, the optimized $T_c$ is achieved only up to 14K[11]

Inset of figure 2 shows the isothermal magnetization $M(H)$ curve of $SmFe_{0.9}Co_{0.1}AsO$ at 2, 5, 7, and 9K. The lower critical field values are seen at around 200, 150, 100 and 50Oe at 2, 5, 7, and 9K respectively. Again confirming the bulk nature of superconductivity in the studied $SmFe_{0.9}Co_{0.1}AsO$. The complete isothermal magnetization loops of the studied $SmFe_{0.9}Co_{0.1}AsO$ at 2 and 5K are shown in figure 3. These plots also point towards the bulk superconductivity of the studied compound. The critical current density roughly estimated from the observed $M(H)$ loops in figure 3 is around $10^3$A/cm$^2$ at 2K and 200Oe applied field.

Figure 4 depicts the isothermal magnetization $M(H)$ plots of the $SmFe_{0.9}Co_{0.1}AsO$ at 50 and 100K, i.e. well above the superconductivity transition temperature of 14K, in applied fields of up to 50KOe. This is to check the possibility of any magnetically ordered impurity phase in the compound particularly in terms of Fe or Co derivatives. As see seen from the two plots at 100 and 50K, the $M(H)$ behavior is linear up till to 5KOe, with only very small remnant moment of around 0.004 emu/g. This value is slightly higher than as observed for the *HPHT* synthesized $SmFeAsO_{0.85}$ ($T_C$ ~52.4K), but still close [14]. This observation rules out any magnetic impurity in our sample. The magnetic moment versus temperature i.e., $M(T)$ plot of the sample in its normal state is exhibited in inset of figure 4. The plot does not follow the Curie-Weiss law and rather show a linear behavior down to at least 100K, later there seems to be an up-turn. The situation is similar to that as in ref.14, with small low $T$ up turn in our sample and slightly higher moment, which may arise due to 10% Co being doped in our sample. The linearity of $Sm^{3+}$ moment can be attributed to Van Vleck ion, being seen in many other Sm compounds as well [15].

In present case the carriers are directly doped within the Fe-As layer via Co doping. Here, the substitution of $Co^{3+}$ at the place of $Fe^{2+}$ in Fe-As superconducting layer provides directly the free mobile electron in this layer and thus suppresses the magnetic *SDW* ordering and stimulate the superconductivity. This is considered as an alternative way to introduce the charge carriers in Fe-As layer without using the fluorine doping or creation of the oxygen vacancies in the charge reservoir Re-O layer. However the transition temperature is low and the volume fraction of superconductivity is also less in comparison to fluorine doped or oxygen deficient samples prepared by *HPHT* technique [14,16-18]. On an *HPHT* synthesized $SmFeAsO_{0.85}$, a $T_c$ of 52.4K is reported with the diamagnetic moment signal at 5K, which is about seven times more than that of the presently studied sample [14]. The less volume fraction in the Co substitution induced superconductor can be attributed to the presence of sufficient disorder produced by $Co^{3+}$ doping at $Fe^{2+}$ site in superconducting layer. Interestingly the same does not hold ground for two-layered $MFe_2As_2$ compound (M = Ca, Sr, Ba or divalent Eu), which exhibits reasonably good superconductivity with Co substitution at Fe site [19]. Perhaps the role of disorder is more prominent in single layered ReFeAsO compounds. Or otherwise the Co substitution induced superconductivity in ReFeAsO [11] is yet far from optimized. As far as the volume fraction and shape of the diamagnetic plots are concerned, our results are close to that as observed in ref. 11 by A. Sefat et al.

In summary, we have successfully synthesized the iron based Co-doped layered compound $SmFe_{1-x}Co_xAsO$ (x = 0 & 0.10) by an easy one step solid-state reaction route. Co doping is an effective way to introduce charge carriers in Fe-As layer to bring out the superconductivity in ReFeAsO system. The challenge between competing direct carrier introduced ($Fe^{2+}/Co^{3+}$) superconductivity and the disorder thus created is yet to be resolved.

## ACKNOWLEDGEMENT

Authors thank Prof. Vikram Kumar Director, National Physical Laboratory for his constant encouragement in present work. Arpita Vajpayee, Monika Mudgel and Anand Pal would like to thank CSIR for financial help in terms of the award of research fellowships to them. Dr. M. Deepa from NPL is acknowledged use of the glove-box. One of the authors VPSA thanks Prof. E. Takayama-Muromachi for the permission to carryout the MPMS magnetization measurements in his laboratory at NIMS Japan and also for various fruitful discussions.

## FIGURE CAPTIONS

Fig 1: Fitted and observed room temperature X-ray diffraction patterns of SmFeAsO and $SmFe_{0.9}Co_{0.1}AsO$ compounds.

Fig 2: Temperature variation of magnetic susceptibility *M*(*T*) in *FC* and *ZFC* condition for studied $SmFe_{0.9}Co_{0.1}AsO$. Inset shows the isothermal magnetization at different temperature for the same.

Fig 3: Complete magnetization loops *M*(*H*) at 2 and 5K for studied $SmFe_{0.9}Co_{0.1}AsO$.

Fig.4 Isothermal magnetization plots *M*(*H*) at 50 and 100K for the studied $SmFe_{0.9}Co_{0.1}AsO$, the inset shows the *M*(*T*) for the same in temperature range of 50 to 250K at 2.5 KOe applied field.

Figure 1.

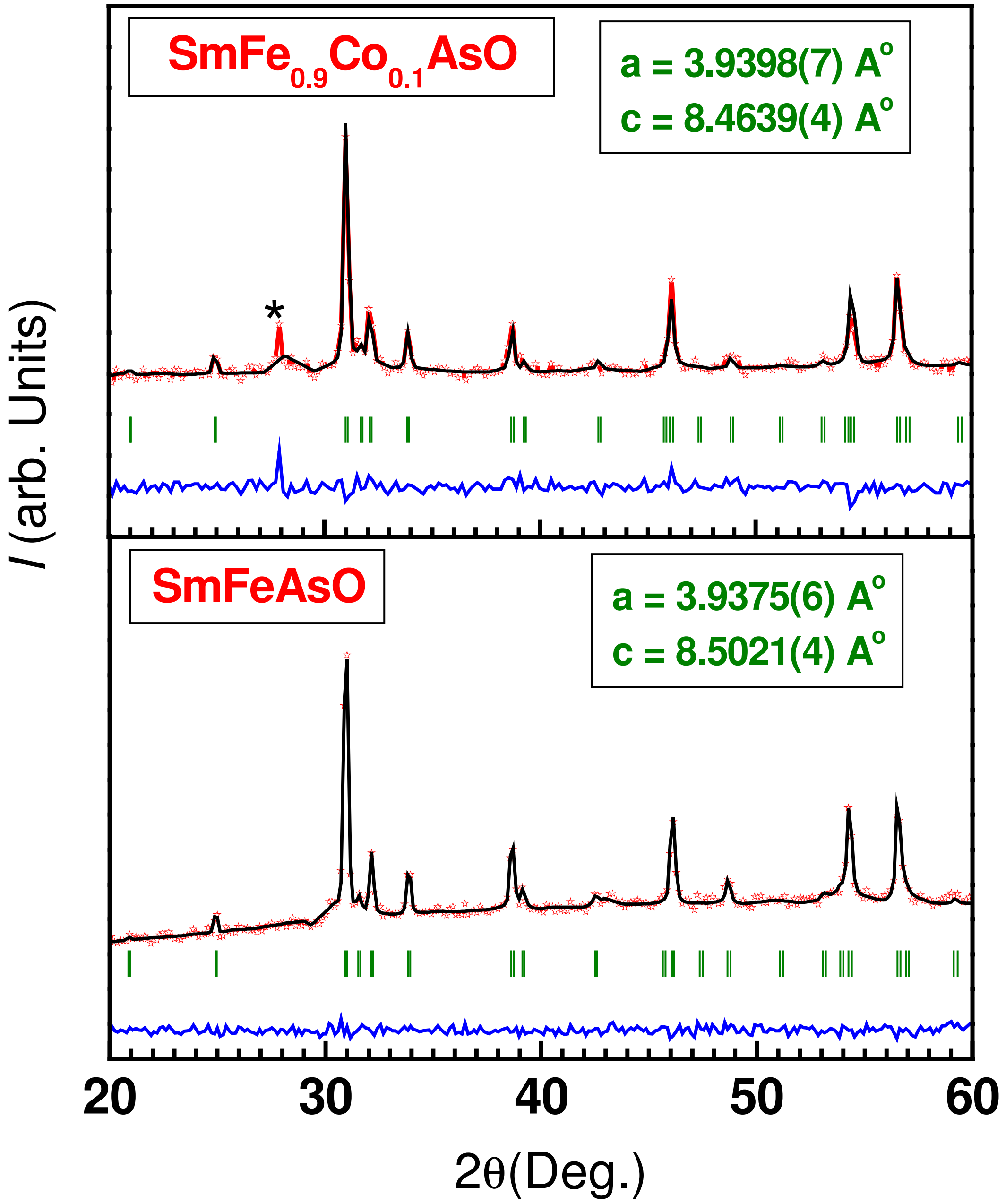

Figure 2.

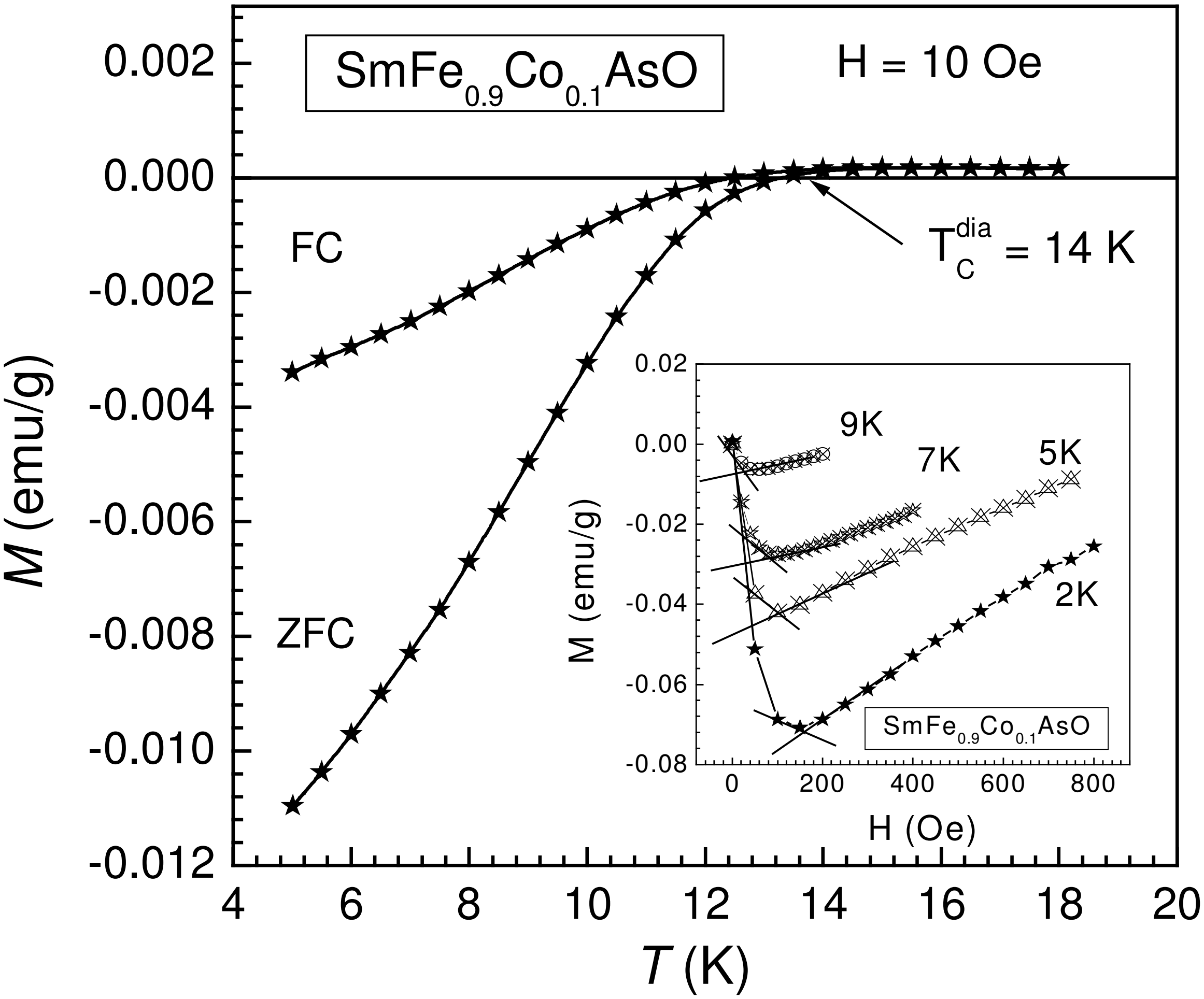

Figure 3.

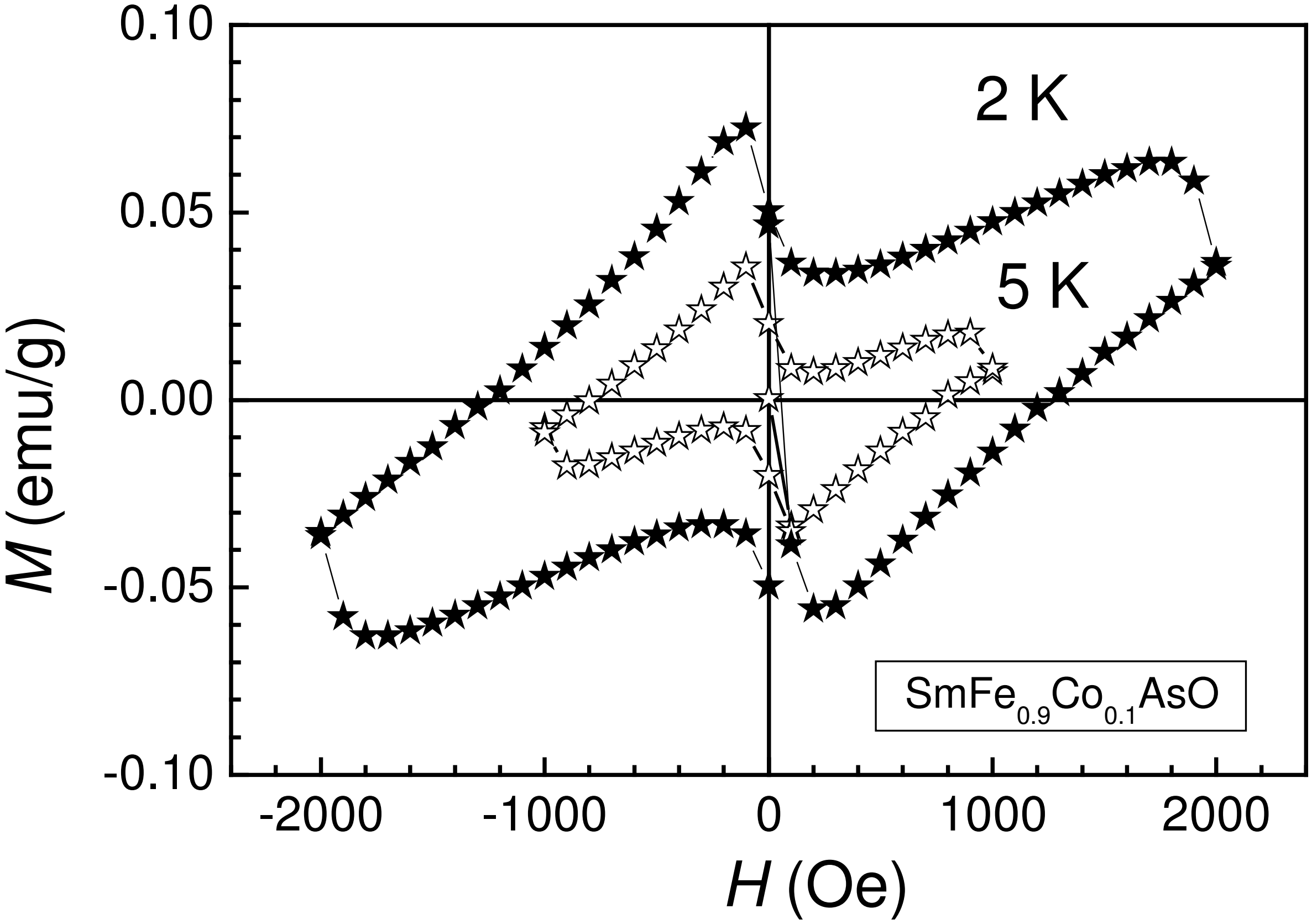

Figure 4

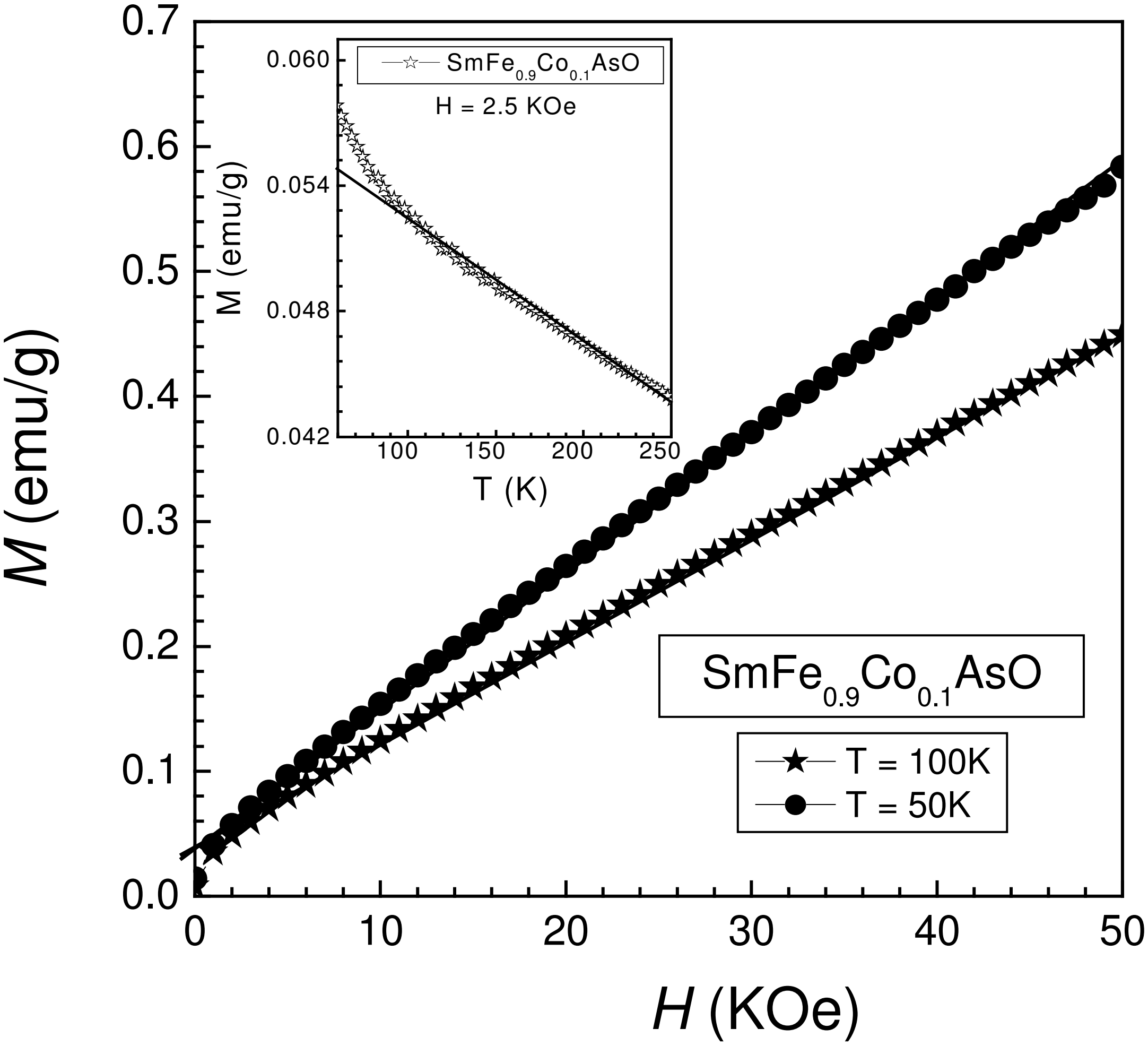